# Energy Consumption Modeling for DED-based Hybrid Additive Manufacturing

Md Rabiul Hasan[a], Zhichao Liu[a] *, Asif Rahman[a]

[a]IMSE, West Virginia University, Morgantown, WV 26501, USA

* Corresponding author. Tel.: +0-000-000-0000 ; fax: +0-000-000-0000. *E-mail address:* zhichao.liu@mail.wvu.edu

**Abstract**

The awareness of energy consumption is gaining much more attention in manufacturing due to its economic and sustainability benefits. An energy consumption model is needed for quantifying the consumption and predicting the impact of various process parameters in manufacturing. This paper aims to develop an energy consumption model for Direct Energy Deposition (DED) based Hybrid Additive Manufacturing (HAM) for an Inconel 718 part. The Specific Energy Consumption (SEC) is used while developing the energy consumption of the product manufacturing lifecycle. This study focuses on the analysis to investigate three significant factors (scanning speed, laser power, and feed rate), their interactions' effects, and whether they have a significant effect.in energy consumption. The results suggest that all the factors have a strong influence, but their interaction effects have a weak influence on the energy consumption for HAM. Among the three process parameters, it is found that laser power has the most significant effect on energy consumption. Again, based on the regression analysis, this study also recommends high scanning speed while the laser power and feed rate should be low. Also, idle time has significant energy consumption during the whole HAM process.

## 1. Introduction

Additive manufacturing (AM) is a newly emerging technology to make three-dimensional (3D) parts directly from CAD model layer upon layer **[1]** . Nowadays, AM is leading a fundamental shift in the way we design and manufacture products **[2]**. It has become the mainstream due to its numerous benefits, including design freedom, material saving, and the reduction of production cost and manufacturing carbon footprint **[3]**. Directed energy deposition (DED) is an AM process in which focused thermal energy is used to melt the deposited materials to make dense 3D structures **[4]**. With good metallurgical bonding, controllable heat input, minimal stress and distortion, cost-effectiveness for remanufacturing, DED is now widely used in high-value components repair **[5]**, prototyping **[6]**, and functionally graded material fabrication **[7]**. HAM integrates AM process with a secondary process, e.g., CNC machining, to additively fabricate a near-net-shape part in a single setup without a need of post-processing. The HAM process is shown in Figure 1, where the additive manufacturing process is used to build the part layer by layer, followed by the machining process to remove the unwanted materials and improve accuracy.

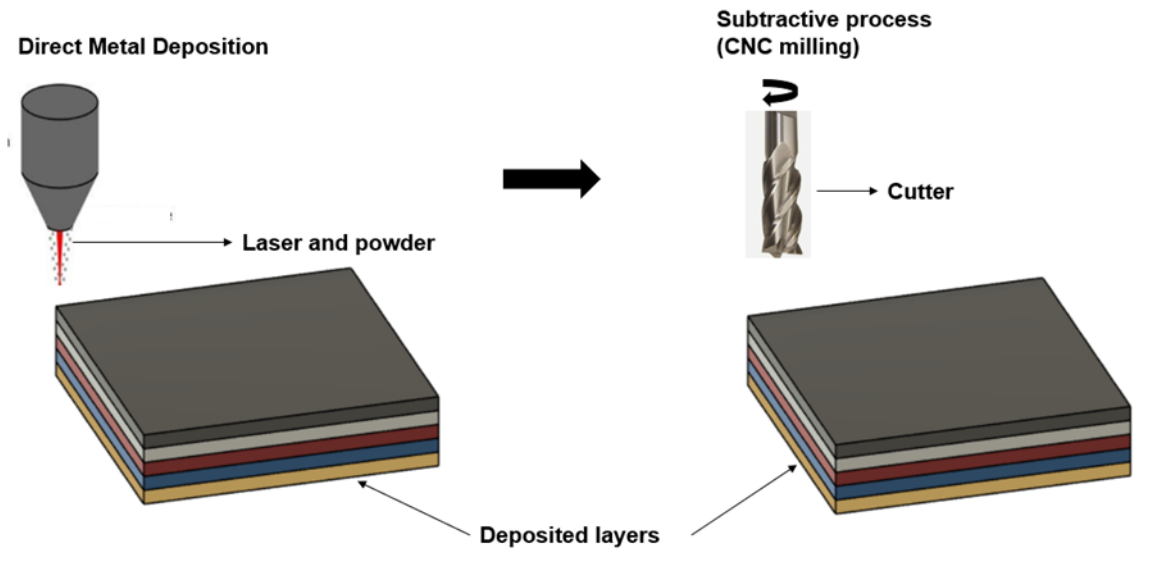

Figure 1: Hybrid Additive manufacturing process (sources from Li et. al., 2020) [8]

Implementation of HAM technology will enhance AM's capabilities in creating high-geometrical-complexity and high-accuracy 3D structures with a smooth surface finish [9]. It will also reduce the cost of post-processing, decrease material waste, and pave way for the subtractive manufacturing industries to enter the fast-growing AM sector [10].

Nowadays, energy consumption is measured as an excellent tool to evaluate the performance of a manufacturing industry. For sustainable development, the target is to reduce global emissions by zero by 2050 [11]. More than 30% of global energy consumption and carbon emissions occur due to manufacturing works [12]. Therefore, there is a trend to investigate and model the energy consumption among different manufacturing processes, and Additive Manufacturing (AM) could not be more exceptional. A recent literature review discovered that most papers calculated the energy consumption for the steady-state processes. They suggested the transient state should be also considered for the energy consumption model scenario [13]. In an energy consumption analysis for AM processes, Gutierrez-Osorio et al. (2019) concluded that there is a strong relationship between layer thickness and energy consumption: a greater layer thickness gives a lower energy consumption [14].

To determine the awareness of energy consumption in manufacturing processes, Owodunni O. (2017) showed that specific energy consumption decreases with an increase in material processing rate for all manufacturing processes [15]. Similarly, Kara S. & Li W. (2011) explored the relationship between MRR and energy consumption. Their study suggested that less energy consumption happens with higher MRR during the unit energy consumption [16]. To understand the energy consumption in wire and powder-based additive-subtractive manufacturing processes, Jackson et al. (2018) found that the wire-based process consumed 24% less energy than powder-based to manufacture a particular object [17]. Recently, machine learning has been applied to predict the power consumption of the machinery tools Darapaneni et al. (2022). They found that the Facebook Prophet algorithm improved more accuracy [18]. Kong et al. (2022) demonstrated that with the integration process planning and scheduling, the energy consumption is less than the non-integrated process planning [19]. Few researchers have been able to draw on any structured research into energy consumption modelling on DED-based Additive-Subtractive Hybrid Manufacturing (ASHM) and the significance of their factor analysis. A recent study found that around 80% of the energy is consumed during the AM for an AHSM process, and they recommended increasing the scanning speed and process rate [20]. Watson & Taminger (2018) found that AM is more energy efficient for volume fractions less than the critical value than subtractive manufacturing, and vice versa [21].

In this study, an energy consumption modeling is built based on specific energy consumption for DED-based hybrid additive manufacturing (HAM) for nickel-based super alloy parts. The effect of the process parameters, including scanning speed, laser power, and feed rate, on the overall energy consumption is analyzed based on Lenth's method. The rest of the paper is organized as follows. Sect. 2 describes the materials and methods, including model development and experiment setup. In Sect. 3, we discussed and explained the results. In Sect. 4, we concluded with an explanation of the result and the future recommendation.

## 2. Materials and methods

### *2.1. Material properties*

As received Nickel-based superalloy Inconel 718 was used as the powder material. The particle size ranges from 44 to 125 µm, and its chemical composition is shown in Table 1. The low carbon steel was used as the substrate material. The carbide tool was used as a cutting tool in CNC milling.

Table 1. Chemical composition of Inconel 718

| Element | C | Cr | Mo | Nb | Ti | Al | Fe | Ni |
|---|---|---|---|---|---|---|---|---|
| wt.% | 0.022 | 18.5 | 3 | 5.2 | 1 | 0.43 | 18.3 | Balance |

### 2.2. Energy consumption analysis

In DED, deposition head movement, laser system, chiller, and fume extractor are significant contributors to the energy consumption. The deposition head and the printing bed move with the help of several servo motors in a non-continuous motion, which consumes energy. The laser machine emits a fiber laser that heats the substrate and melts the powder. The laser is turned off after each layer is printed to prevent the accumulation of excess molten materials. The chiller is operated to protect the optic and laser system. The fume extractor is turned on after one part is printed to extract the hazardous gas generated during fabrication. All these components require a non-continuous power supply.

In CNC, the material is removed from the workpiece with a cutting edge in the form of chips. Energy consumption in the CNC process depends on the material removal rate, feed rate, spindle speed, and depth of cut. Coolant can be used to reduce the cutting force hence increasing the tool life and reducing the energy consumption. During the CNC, energy is consumed for the operation of motorized spindles and other positioning equipment and the metal removal process. The energy consumption during processing also depends on the machine's production rate and the material being processed. Energy will also be consumed during idle time in both DED and CNC.

### 2.3. Model development

Specific Energy Consumption (SEC) is used to develop an energy consumption model in HAM. SEC is defined as the energy consumed in the production of a material unit, which depends on the process characteristics. In CNC, SEC is the energy consumption per unit volume of material removal while in DED, SEC can be determined as energy consumption per unit mass of material deposition [21].

For HAM processes, it is critical to establish a common framework that can describe all energy inputs on a standard basis. Energy consumption in each stage of HAM must be included in the calculation of the energy consumption modeling. These quantities account for the total energy associated with DED and CNC. Thus, an expression for the energy required for hybrid additive manufacturing can be defined as in equation (1):

$$E_D = \alpha V_T E_{VD} + f \alpha V_T E_{VM} \qquad (1)$$

Where,

$V_M$ = volume of material deposited in the additive manufacturing

$V_T$ = Volume of material defined in the part design

$\alpha$ = Fraction of part design containing solid material “solid to design ratio” ($V_M / V_T$)

$E_V$ = Energy consumption during the material deposition in additive manufacturing

$V_R$ = Volume of the materials removed in subtractive manufacturing

$f$ = Fraction of deposited materials removed by machining

$E_{VD}$ = Energy consumption in unit volume of material deposition in additive manufacturing ($E_V / V_M$)

$E_{VM}$ = Energy consumption in unit volume of material removal in subtractive manufacturing ($E_S / V_R$)

$\alpha V_T E_{VD}$ = Specific energy consumption for additive manufacturing

$f \alpha V_T E_{VM}$ = Specific energy consumption for subtractive manufacturing

EVD and EVM is the energy required for deposition and machining. They not only include the direct energy consumed by the part fabrication or the machining process but also the indirect energy consumed by the processing equipment, such as start-up, repositioning, or tool changing time. These nonproductive components of energy consumption can be substantial and equal to or greater than the energy consumed by the actual material processing action. Values for EVM and EVD are material-specific and machine-specific. They can be estimated by experimentally measuring the energy consumed per unit volume of each material that might be processed. Non-machining time or idle time is not considered while calculating energy consumption modeling. The data loggers capture the idle power of 2.30 kW for the DED of AM process.

*2.4. Experimental setup*

In this study, a customized hybrid manufacturing system (AMBITTM, Hybrid Manufacturing Technologies, Texas, USA), is used to fabricate/machine the part. A laser deposition module is integrated with HAAS TM-1P milling machine, as shown in Figure 2. The AMBITTM hybrid system consists of a 1000 W IPG fiber laser system, a pneumatic powder feeder system, and a computer-controlled motion system. The powder is delivered to the melt pool created by a highly focused laser beam. Argon is used as the carrier and shield gas. The travel path is controlled by the CNC control panel to fabricate the designed object. Both ABMIT core and CNC machine are connected and operated in a 3-phase 220V condition to carry out the hybrid operations. The current transducer and data logger has been used to measure the energy consumption during machining. The time interval between each data point is set as 1 sec. The power factor of the machines is considered 85 percent.

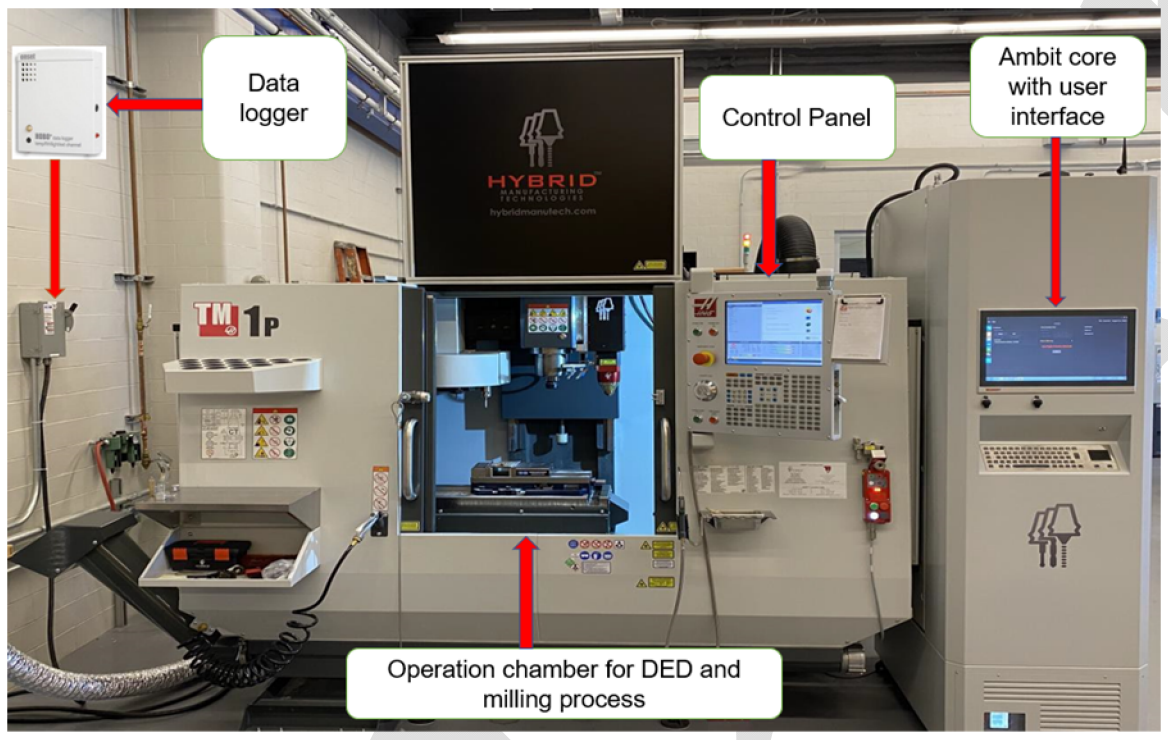

Figure 2: Machine setup of HAM

The CAD design of the part to be printed is shown in Figure 3. The length and width of the part are 15 mm each. For additives, five layers are designed to be printed with a thickness of 0.54 mm for each layer. 0° and 90° rectilinear scanning strategies were implemented. The total thickness of the part is 2.7 mm. The process parameters for the experimental design of the HAM process are listed in table 2. The other process parameters of the HAM process are shown in Table 3.

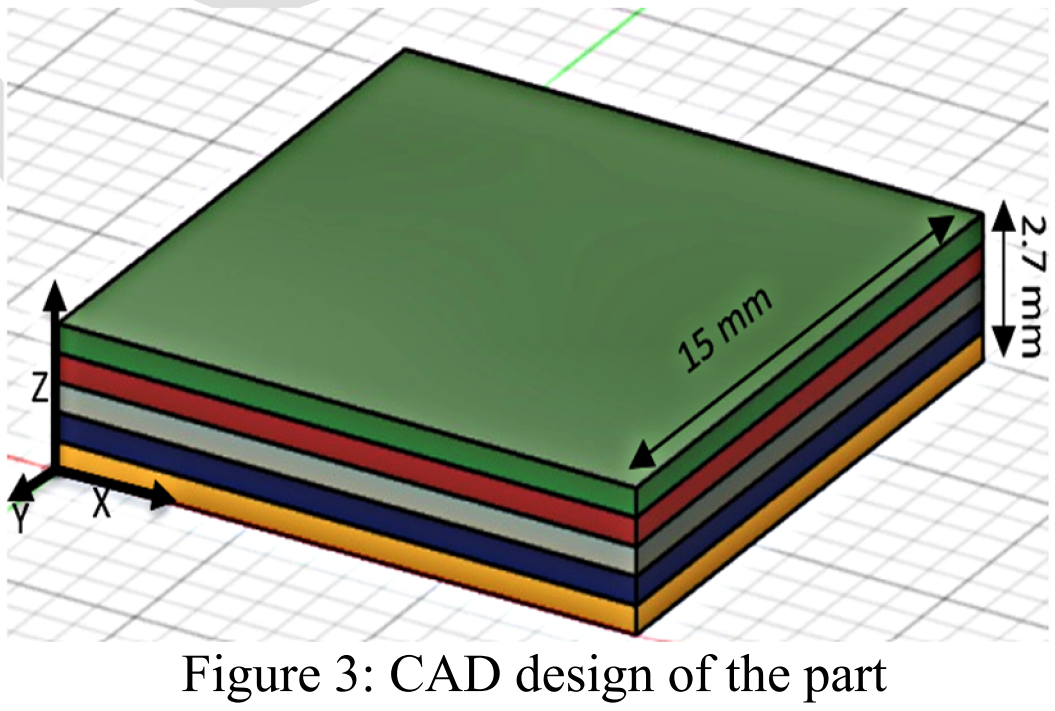

Figure 3: CAD design of the part

Table 2: The process parameters for experimental design of HAM process

| Experiment No | Scanning Speed (mm/s) | Laser Power (W) | Feed Rate (rpm) |
|---|---|---|---|
| 1 | 600 | 700 | 0.3 |
| 2 | 600 | 700 | 0.5 |
| 3 | 600 | 900 | 0.3 |
| 4 | 600 | 900 | 0.5 |
| 5 | 800 | 700 | 0.3 |
| 6 | 800 | 700 | 0.5 |
| 7 | 800 | 900 | 0.3 |
| 8 | 800 | 900 | 0.5 |

Table 3: Experimental setup

| Process parameter in DED | | Process parameter in CNC | | |
|---|---|---|---|---|
| Nozzle gas (I/min) | Shield gas (I/min) | Spindle Speed (rpm) | Spindle Power (kW) | Feed rate (MMPM) |
| 9 | 12 | 1526 | 0.6 | 155.2 |

## 3. Results and Discussion

### *3.1. Energy consumption profile*

Totally, eight blocks were fabricated with HAM process based on the parameters in Table 3, as shown in Figure 4. For the first part, both DED and CNC are performed. For the other seven parts only, DED based AM is performed.

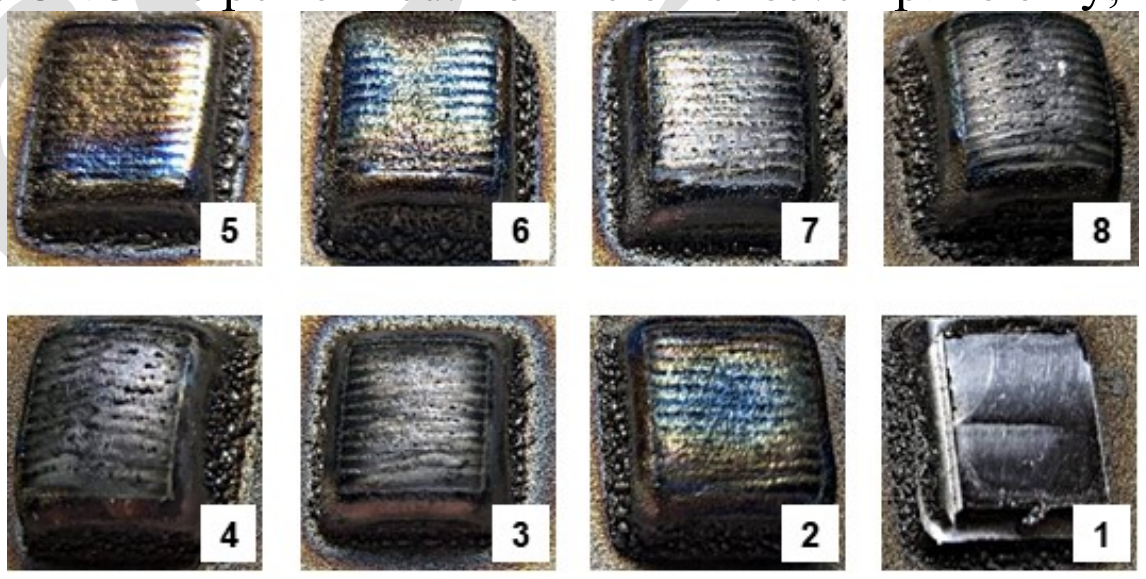

Figure 4: Final parts manufactured by HAM process

The current consumption profile for DED AM and CNC machines is plotted in Figures 5 and 6. Eight spikes in Figure 5 represent the energy consumption profile for the eight parts printed in the DED process. Each spike is divided into five smaller sections representing the five layers in the printed part. It is evident that laser power has higher impact on energy consumption compared with scanning speed and federate. When the laser power is 700 W, the current is about 15 amps for part 1, 2, 5 & 6. When the laser power is 900 W, the current is about 17 amps for part 3, 4, 7 & 8. The current during idle time is about 8 amps, and this is mainly consumed by chiller, motion control and fume extractor systems. In Figure 6, the spikes represent the energy consumption in the facing and contouring process in CNC. The energy consumption in CNC is insignificant compared to DED AM. Since the cutting conditions in CNC remain the same for all the parts, energy consumption in the CNC process is assumed to be similar, therefore. CNC is only performed in the first part.

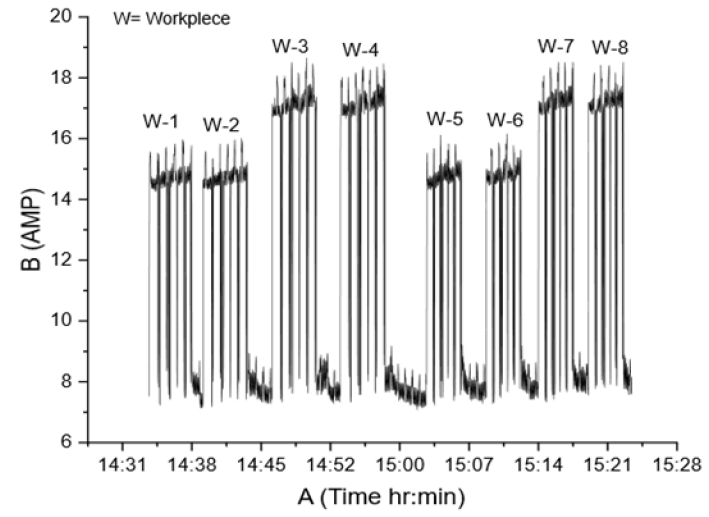

Figure 5: Current consumption profile during additive manufacturing of the part

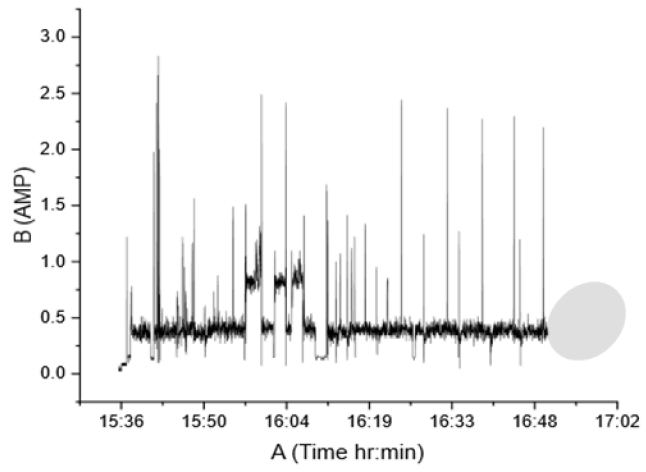

Figure 6: Current consumption profile during subtractive manufacturing of the part

Energy consumption for DED and CNC was measured with the data logger, and the result is listed in Table 4. Specific energy consumption in HAM is calculated using the mathematical model developed in equation 1.

Table 4: Energy consumption in AM and SM

| No of experiment | Energy consumption in AM, Ev (kWh) | Energy consumption in SM, Es (kWh) | Specific Energy Consumption in HAM, ED, (kWh) |
|---|---|---|---|
| | | | |
| 1 | 0.37 | 0.04 | 1.14 |
| 2 | 0.36 | 0.04 | 1.2 |
| 3 | 0.42 | 0.04 | 1.31 |
| 4 | 0.41 | 0.04 | 1.36 |
| 5 | 0.35 | 0.04 | 1.07 |
| 6 | 0.34 | 0.04 | 1.11 |
| 7 | 0.34 | 0.04 | 1.19 |
| 8 | 0.32 | 0.04 | 1.21 |

The distribution of energy consumption in HAM processes is shown in Figure 7. 61% of the energy is consumed for the DED process, whereas 6% of the energy is consumed in the following CNC process. 32% of the energy is consumed when the DED machine is idle. Machining and idle time are shown in Figure 8. 15% of the time is consumed in the DED process, whereas 4% of the time is consumed in the CNC process. The rest is idle time.

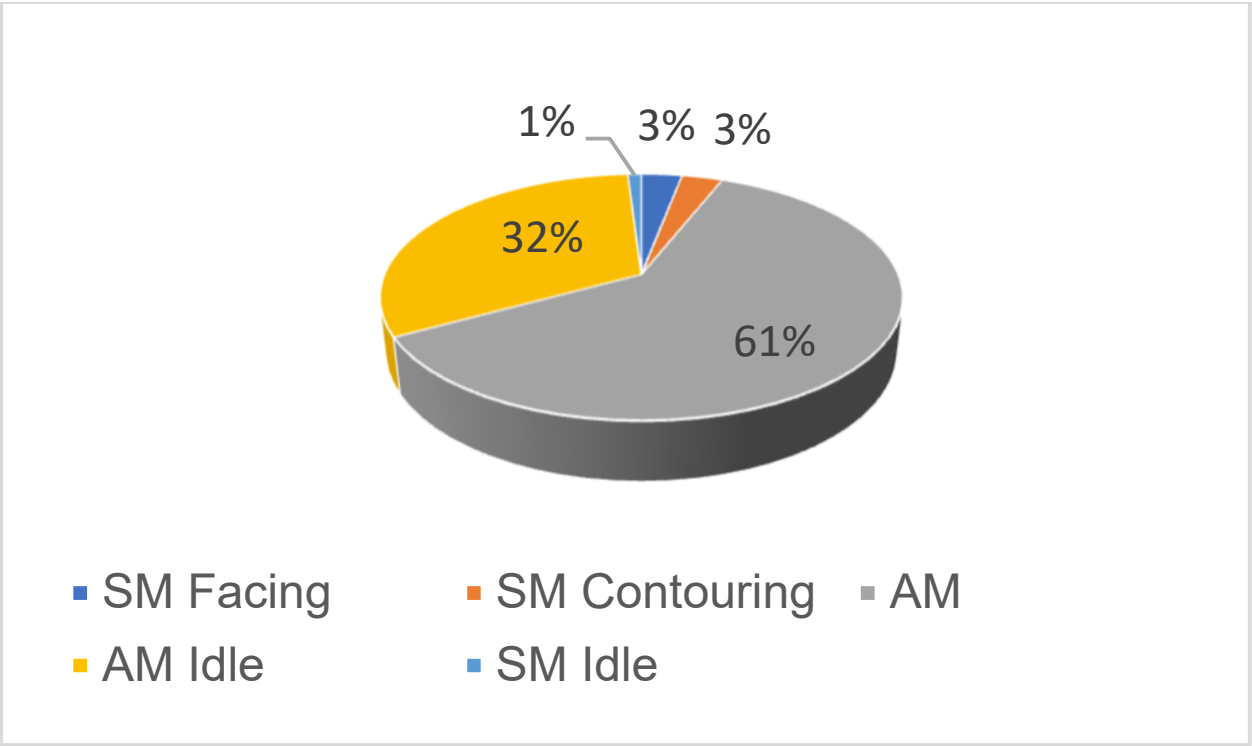

Figure 7: The distribution of energy consumption of HAM process

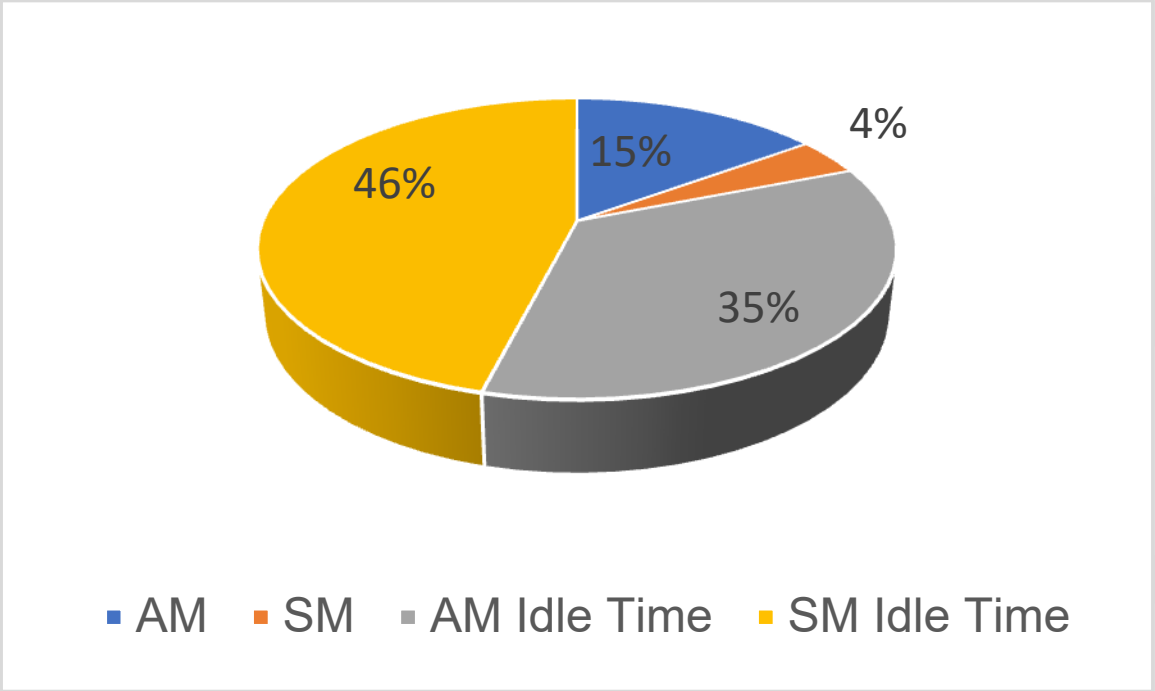

Figure 8: Machining and idle time of AM and SM process

### 3.2. *Effect of process parameters on energy consumption*

The effect of process parameters on the overall energy consumption is analyzed with Length's Method [22]. The complete factorial analysis is conducted at each combination of the primary and interaction effects levels. Due to resource and time constraints, a single experiment is carried out without replication in this study. Applying Lenth's method, the graphical approach is used to construct Half Normal plots of the estimated effects and identify "significant" and "not significant." Determining the "significant" effects will lead to an estimate of standard deviation (σ), the remaining effects are responsible for the chance variation.

With this method, we considered a robust estimator of the standard deviation of Median ($\hat{\theta}_i$), which is called the pseudo standard error or PSE. A robust estimator means its performance is not sensitive to the ($\hat{\theta}_i$) associated with dynamic effects [22].

$$PSE = 1.5 \text{ X Median } (\hat{\theta}_i) \qquad (2)$$

In this formula Median is used to provide robustness against outliers. Median in the formula of PSE is computed those ($\hat{\theta}_i$) satisfying that.

$$(\hat{\theta}_i) < 2.5\, S_o. \qquad (3)$$

$$\text{Where, } S_o = 1.5 \text{ X Median } (\hat{\theta}_i) \qquad (4)$$

In equation (1) the initial standard error $S_o$, scaling factor 1.5 is used. It is a consistent estimator of the standard deviation of $\hat{\theta}_i$ when the $\hat{\theta}_i$ is zero and the underlying error distribution is normal [22]. After that $t_{PSE}$ is considered. $t_{PSE}$ can be found by dividing the interaction effect of each factor by PSE. Therefore,

$$t_{PSE,i} = \frac{\hat{\theta}_i}{PSE} \text{ for each } i \qquad (5)$$

If | $t_{PSE,i}$ | exceeds the critical value, then i is declared as significant.

Two versions of critical values are generally used. They are: Individual Error Rate (IER) and Experiment Wise Error Rate (EER). Now, if $H_0$ is characterized as the null hypothesis where at the level $\alpha$ , IER$\alpha$ and EER$\alpha$ stand for the IER and EER based critical values of $t_{PSE}$ , then

$$\text{Prb } (|t_{PSE,i}| > \text{IER}_{\alpha} \mid H_0) = \alpha \qquad (6)$$

Where I =1,….*I,* and Prb ($|t_{PSE,i}| > \text{EER}_{\alpha}$ for at least one i, I = 1,….,*I*| $H_0$) = Prb ($\max_{1 \le i \le I} t \, |t_{PSE,i}| > \text{EER}\alpha \mid H_0) = \alpha$.

EER is measured according to the number of tests done in the experiment. However, it is used less as it gives conservative results and is less powerful. Due to the power to identify the significant effects, the IER version of Lenth's method is more recommended than the EER-based Lenth's method [22].

In this paper, three process parameters have been considered, including Scanning Speed (X), unit: mm/s, Laser Power (Y), unit: Watt (W) and Feed Rate (Z), and unit: Rotation per minute (rpm). Each factor has two levels – High and Low. The factors and levels are described in the Table 5 below:

**Table 5:** Process parameters and their levels

| Parameters | Level 1 (Low) | Level 2 (High) |
|---|---|---|
| Scanning Speed (X) (mm/s) | 600 (-) | 800 (+) |
| Laser Power (Y) (W) | 700 (-) | 900 (+) |
| Feed Rate (Z) (rpm) | 30%(-) | 50%(+) |

The number of effects includes main and interaction effects: 2k – 1. Here, k = 3. therefore, the total number of effects is 23 – 1 = 7, which are X, Y, Z, XY, YZ, XZ, and XYZ. Minitab software is used to carry out the factor analysis. Based on the required energy model developed in equation 1, the total energy for each combination of process parameters in the regression equation for this experiment design has been shown below, referred to as Total energy (ET) consumed during HAM process:

ET = 4.310 - 0.004654 Speed
+ 0.000852 Power+ 0.03085 Feed+ 0.000006 Speed*Power- 0.000006 Speed*Feed- 0.000002 Power*Feed- 0.0000 00 Speed*Power*Feed (7)

The design matrix and the output energy consumption according to the developed energy consumption model in equation (7) has been shown in Table 6.

**Table 6:** Design matrix

| Run | X | Y | Z | XY | YZ | ZX | XYZ | ET (KW) |
|---|---|---|---|---|---|---|---|---|
| 1 | - | - | - | + | + | + | - | 4.88 |
| 2 | - | - | + | + | - | - | + | 5.154606 |
| 3 | - | + | - | - | + | - | + | 5.609072 |
| 4 | - | + | + | - | - | + | - | 5.807712 |
| 5 | + | - | - | - | - | + | + | 4.578857 |
| 6 | + | - | + | - | + | - | - | 4.750211 |
| 7 | + | + | - | + | - | - | - | 5.498789 |
| 8 | + | + | + | + | + | + | + | 5.571965 |

For this analysis, α=0.01 is considered. When I=7, the IER is 5.07. Comparing with 2.5So (2.5So = 337.039) it is found that the value of the main effect Y (Laser Power) is bigger (780.966). Therefore, Laser Power is significantly responsible for the most energy consumption. Other than factor Y, no other factors significantly affect energy consumption according to Lenth's method. Then, further analysis is conducted with half normal plot to identify the effect.

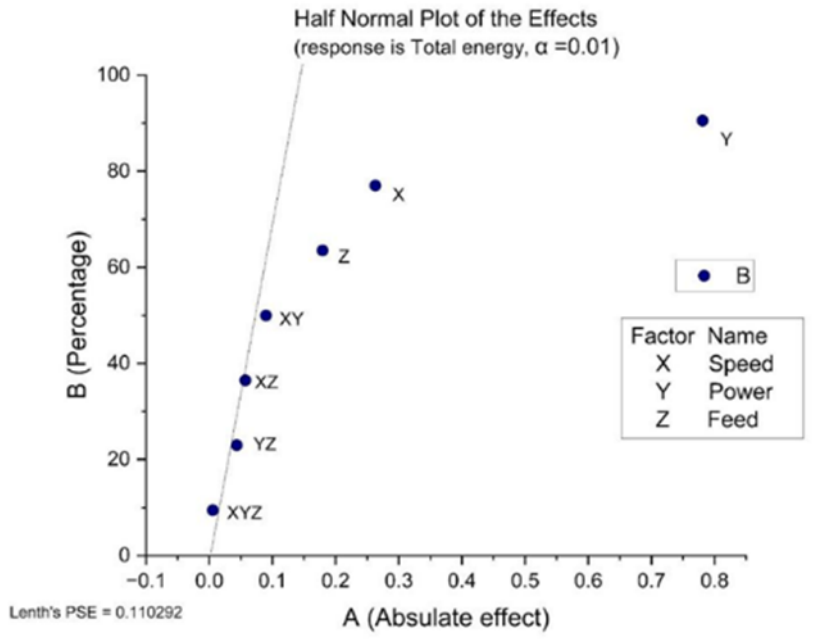

Figure 9: half normal plot

The half normal plot representing the effect of the factors in shown in Figure 9, indicating that Y , X and Z significantly affect the energy consumption during the HAM process. However, other interaction effects of XYZ, YZ, XZ, and XY are insignificant as they are not departed from the linear line. Therefore, from the regression equation, the laser power and the feed rate should be decreased while the scanning speed should be increased to minimize the total energy consumption of the HAM. This result is reflected in the fifth experimental run described in table 5 with the lowest energy consumption (4.579 KW) and the referred combinations of significant factors.

## 4. Limitations and Future works

### *4.1. Limitations*

The part processing energy depends on which processing parameters are used during manufacturing. Due to the cost and time limitation, this study only investigated the three primary parameters and their interaction effects on the energy consumption model. The reader should bear in mind that this research is based on the full factorial design in regression analysis, which is considered for the experiment parts. It can only be possible when a number of experiments have been run. It generates the problem when the parameters exceed seven or more ( k⩾7). So, the full factorial design will become more difficult to run from an economic perspective. Other techniques like response surface methodology, the fraction of full factorial design, or robust factor design can be other choices to explore for this research with more design parameters. The energy model which is developed here depends on the customized HAM machine. It needs to add other energy sources based on the individual processing machines, which consume energy while producing a part; in post-processing and return of post-processing scrap material for recycling.

### *4.2. Future works*

The energy consumption of HAM has its advantage due to its manufacturing efficiency, and further work is required to establish this. Some of the potential future works which are not limited are enlisted below:

- Different machine learning algorithms can be used to predict and analyze the HAM process.
- The effect of different materials composition can be considered for the energy consumption modeling.
- There is an opportunity to extend this research work by analyzing other essential parameters in HAM.
- In future research, energy consumption models can be developed by considering the transient state of the manufacturing process.
- Energy consumption models can be explored based on different part designs.
- There is room for further progress to compare energy efficiencies among different HAM processes.

## 5. Conclusions

In this study, the energy consumption model has been developed for the HAM process. Energy consumption has been measured both for the AM and SM processes. Specific energy consumption has been calculated using that mathematical model. It has been found that the energy consumption of CNC machining is very low compared to the AM of the HAM. It is pointed out that the processing parameters of AM and their effects are of more concern that should be investigated than the processing parameters of SM in HAM. So, the effect of three significant parameters (scanning speed, laser power, feed rate) and their interaction effects in AM process has been investigated. From this study, it is discovered that all the main factors significantly contribute to the energy consumption of the HAM process. However, among the three parameters, the laser power (Y) effect is the highest. Moreover, for the reduction of the energy consumption of DED-based HAM, it is recommended that the scanning speed (X) should be higher while both the laser power (Y) and feed rate (Z) should be lower. So, the results are significant for modeling and reducing the energy consumption of the HAM process, which provides further support to investigate the energy consumption of HAM in the upcoming days.